\documentclass[12pt]{article}
\batchmode

\usepackage{graphicx}

\begin{document}

\date{}

\title{Routing Packet Traffic via Enhanced Access Control List for Network Congestion Avoidance}

\author{
\Large  \hspace*{7mm} Zohre R. Mojaveri and Andr\'as Farag\'o\\
Department of Electrical and Computer Science \\
The University of Texas at Dallas \\
Richardson, Texas\\
}

\maketitle
       \thispagestyle{empty}

\noindent
{\bf Abstract --
Filtering packet traffic and rules of permit/denial of data packets into network nodes are granted by facilitating Access Control Lists (ACL). This paper proposes a procedure of adding a link load threshold value to the access control list rules option, which acts on the basis of threshold value. The ultimate goal of this enhanced ACL is to avoid congestion in targeted subnetworks.  The link load threshold value allows to decide that packet traffic is rerouted by the router to avoid congestion, or packet drop happens on the basis of packet priorities. The packet rerouting in case of high traffic loads, based on new packet filtering procedure for congestion avoidance, will result in the reduction of the overall packet drop ratio, and of over-subscription in congested subnetworks. 
}


\newpage
\section{Introduction}

Current massive volume of packet traffic in telecommunication networks require smart strategies for handling unforeseen situation under control. 
Nodes in computer networks and specifically ports in network devices may experience heavy packet traffic loads at any given time as a result of situational circumstances of network usage.  
Network administrators attempt to set policies by using available tools to facilitate best possible inbound and outbound packet traffics for the network nodes under their control. Facilitating data transfer between network ports or data processing systems in a network that utilize a common link in the network require planning in order to avoid congestion. Packet oversubscription happens when packet traffic saturation happens in subnetworks. Rerouting packet traffic from heavily loaded ports and links in a subnetwork into lightly loaded links and ports could be an approach that contributes to avoiding congestion in a subnetwork specifically when the system desires or promises to prevent packet drop and oversubscription.  

When network nodes are heavily overloaded in a subnetwork and ports cannot keep up with packet rates, congestion will occurs. Utilizing traffic load parameters and congestion feedback for flow rate adjustment mechanism and rerouting or allocation of extra resources in network flow setting, could decrease delay and packet loss in targeted network and consequently less oversubscription packets or delayed time stamped packets. There are researches about ACL reconstruction framework that suggest replacing an inefficient and redundant Access Control List rule list into a completely new rule list based on traffic tendency\cite{waka}. Hence suggesting enhancements to ACLs as current methods of packet filtering in networks, could increase overall performance of network components specifically routers. 

The organization of this paper is as follows. In next 
section we present some details about Access Control List. In Section three we give a brief survey of Congestion Avoidance Routing. In Section four the proposed Enhanced Access Control List for Network Congestion Avoidance is presented. In Section five results of packet traffic simulation are shown and finally a summary section is presented.

\section{Access Control Lists}
One of the most fundamental component for packet traffic filtering in computer networks is access control lists (ACL). Packet filtering or ACL controls inbound or outbound packet traffic and provides the ability to manage network traffic flowing through a networking element to optimize quality of service (QoS),  network security and performance. Normally determination of permitting and/or denying of a packet is network administrator’s responsibility and these rules are manually configured.  Each ACL contains at least one or more separate action sets which are the defined rules that filters network traffic by controlling whether packets are forwarded or blocked based on each specific criteria in ACL rules. 
Following section covers ACL’s rules and specifications for an active network node packet analyzer tool platform:

\begin{itemize}
\item User construct all rules of an ACL in a sequential order.

\item Rules sequences should be defined from very specific to very general to avoid dropping desired packets. 

\item Within a single rule, all specified fields form an AND operation.

\item To create an OR operation separate rule must be created.

\item Network packet analyzer tool evaluates all rules of ACL in a single clock cycle.

\item More than 30 rule lines can be included in a single ACL.
\item ACL rules are part of stacks that applies to traffic flows.

\end{itemize}

Each ACL contains sequence number for each rule. Rules actions are permit or deny. ACL filtering is based on Ethertype, IP Protocol, Packet Length, VLAN – (Inner or Outer), Source IP and MAC, Ingress Port and Destination IP and MAC address. 

Filtering of packet traffic can be applied for raw Ethernet packets, IPv4 incoming packets or IPv6 incoming packets. 
In general ACL act as filters for network traffic, packet streams and services so this easily filtering and selectively permit or deny network traffic by individual ingress or egress ports or to an entire ingress or egress traffic flow need to be considered carefully in regards to interface ingress or egress this should apply to satisfy the purpose of ACL. Selectively permitting or denying traffic based on specific criteria is supported as follow by ACL considering different network layers:
\begin{itemize}
\item Layer 2: MAC,VLAN,MPLS or Ethertype
\item Layer 3: Source and destination IPv4 and IPv6 sessions, DSCP or IP protocol. 
\item Layer 4: Port number or TCP Control
\end{itemize}

Matching criteria can be on single entry or a range of values of MAC addresses and IP addresses. Masking provides mechanism for comparing range of values and all MAC, IPv4 and IPv6 addresses can be masked.
Creating, Editing and deleting ACLs via CLI or GUI:
When creating ACLs entries lines are added one at a time in a sequential order. ACLs can be reordered, edit or delete via Command Line Interface (CLI) or Graphical User Interface (GUI). It is important to apply created ACL to correct interface and decide where at a given port would be best to apply ACL when it comes to select ingress or egress side of deployed network flows. 

There are also rate-limiting ACLs in routers like in Cisco routers. For example the three rate functions can be defined in router configuration. 
First the average rate, specified in bits per second (bps), for the matching traffic. This is measured by a long-term average of the transmitted rate of traffic on the interface. Traffic under this rate is considered to be conforming.
Second the normal burst size, specified in bits per second (bps). This determines how long traffic can burst above the average rate before it is considered nonconforming and third the excessive burst rate, specified in bits per second (bps). Traffic that exceeds the excessive burst rate is considered nonconforming\cite{cisco}.

\section{Congestion Avoidance Routing via Enhanced ACL}

Routing packet traffic in an optimal approach that considers resource limitation is a key element in congestion avoidance techniques for telecommunication networks. Congestion control monitors the network for conditions that could potentially degrade performance when the system is under heavy load. Typically these condition are temporary for example this scenario of so many mobile users attempt to access network for uploading large types of file like videos in an stadium watching a game. Detecting overloaded links in the network and redistribute packet traffic load to other possible routes in the cluster at appropriate time leads to a better overall performance of network. Dense subnetworks in above mentioned example can be identified using different mathematical approaches and techniques \cite{farago} therefore redistributing packet traffic becomes feasible action for heavy packet traffic load occasions if routers have the ability avoiding congested links in targeted subnetworks.

Congestion control operation is based on configuring the following thresholds in targeted subnetwork.
\begin{itemize}

\item Subnetwork Congestion Condition: Thresholds dictate the conditions for which congestion control is enabled and established limits for defining overall state of the system that means deciding if subnetwork is congested or clear.  

\item Link Utilization: Setting link utilization threshold, when the average utilization of all links in subnetwork reaches the specific threshold, a congestion control mechanism shall be enabled. This paper only focus on link utilization threshold. 

\item Port Utilization: Ingress or egress port specific thresholds can be set and when threshold is reached, congestion control shall be enabled for the nodes in targeted subnetwork.

\end{itemize}

\begin{figure}[ht]
\centering
\hspace*{-5mm}
\includegraphics[width=0.8\columnwidth]{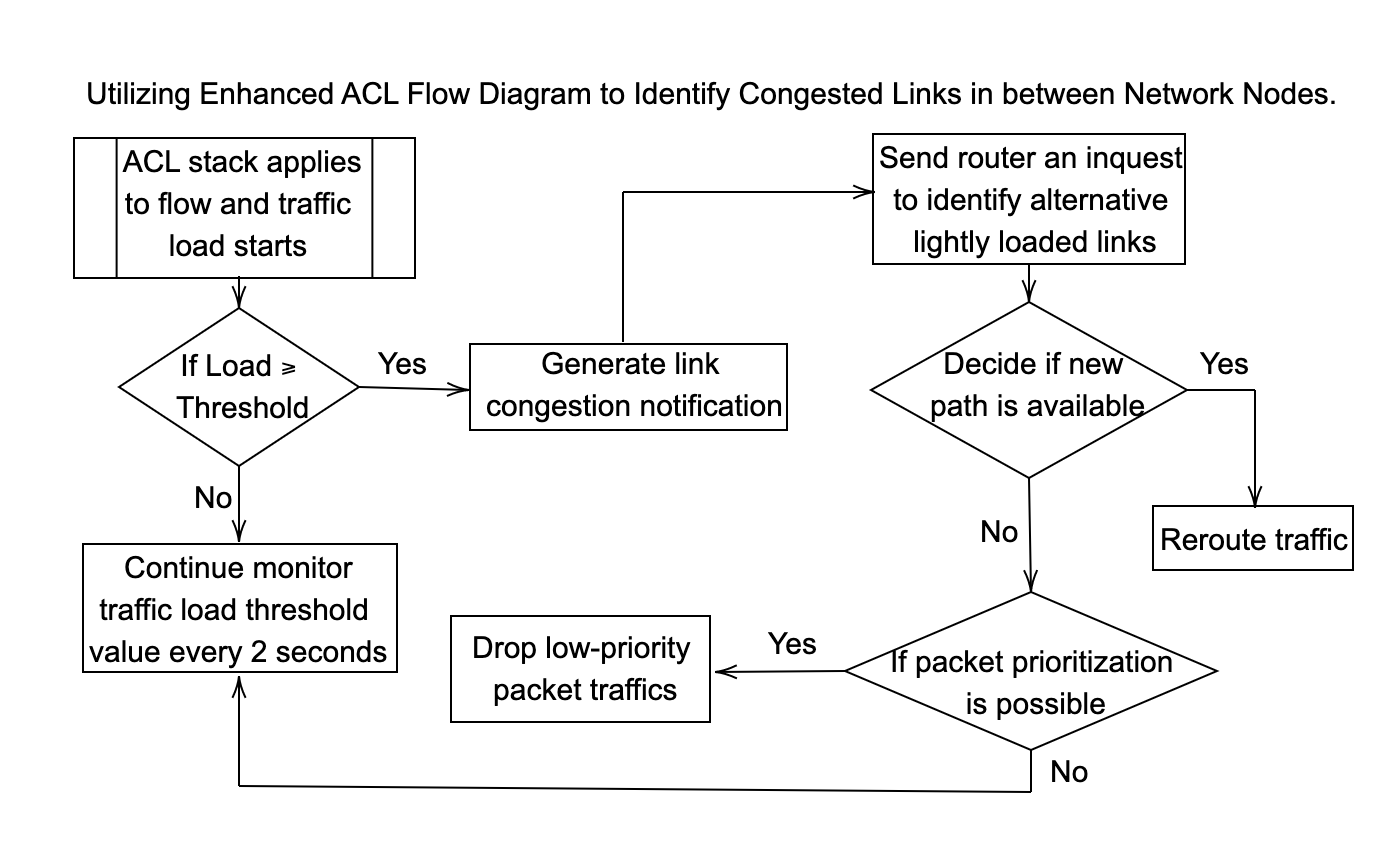}
\caption{Algorithm Flow Diagram}
\label{img:flow diagram}
\end{figure}

\section{Algorithm Development of Enhanced ACL}
Proposed algorithm covers cases of actions to take when there exist traffic load more than the threshold specified for selected links in targeted subnetworks. The algorithm considers that a specific ACL stack is applied to ingress port and it's associated link. 
Algorithm flow diagram is shown in Fig.~\ref{img:flow diagram}.
Our algorithm attempts to consider links between ports and their line rate capacity as cause of possible congestion. Link weight can be derived from various practical parameters that are related to the load in between connected ports of each subnetwork. Proposed ACL adds a filtering parameter based on momentary packet traffic load calculation and initiates an alert when packet traffic load exceeds pre-specified thresholds. This alert generation and filtering feature may work as a base for rerouting packet traffic loads at given times that system is under heavy data transferring situations. For future work dropping packet based on their specified priority and IP protocols can becomes as an active action if rerouting is not a feasible option at that moment.

\section{Packet Traffic Simulation}
\subsection{Modeling overloaded ports of network nodes}
Overloaded links in communication networks is a challenge that requires consideration when designing related architecture and requirements of network to achieve routing of packet traffic from sources to their destinations with none to very limited packet drop counts. Packet frame loss happens when the percentage of frames that should have been forwarded by a network node under steady state load, were not forwarded due to lack of resources and this problem is very probable to happen when the packet traffic load hitting more than 90 percent of intended packet traffic load.

\begin{figure}[ht]
\centering
\hspace*{-5mm}
\includegraphics[width=0.8\columnwidth]{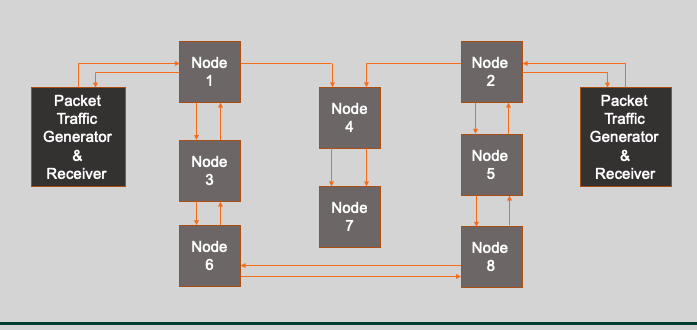}
\caption{Network Nodes and Traffic Generators/Receivers}
\label{img:10Nodes}
\end{figure}

\begin{figure}[ht]
\centering
\hspace*{-5mm}
\includegraphics[width=0.8\columnwidth]{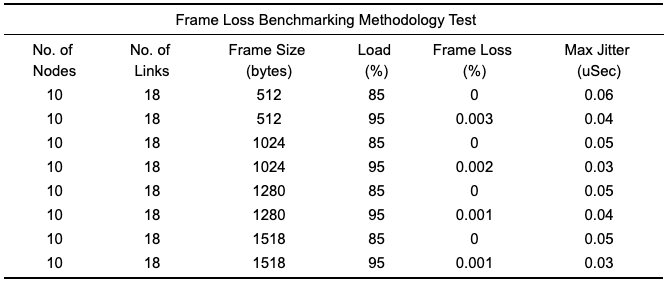}
\caption{Frame Loss Benchmarking Methodology}
\label{img:Frame Loss}
\end{figure}

\begin{figure}[ht]
\centering
\hspace*{-5mm}
\includegraphics[width=0.8\columnwidth]{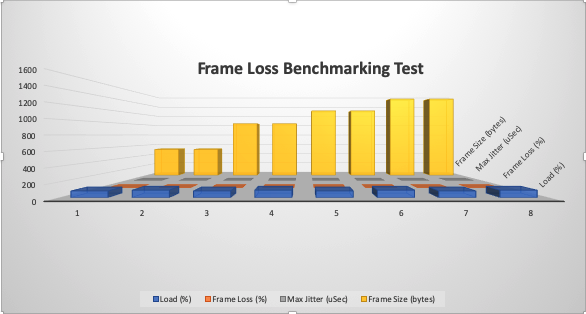}
\caption{Frame Loss}
\label{img:Frame_Loss}
\end{figure}

The following section covers a simulation experiment model of overloaded links in between ports of network nodes. In our simulation modeling overloaded ports of network nodes and correspondents links in between them simulated with various packet traffic patterns. 
Simulation is based on benchmarking methodology for network interconnect devices. The following test scheduling has FIFO (Bit forwarding) latency type and scheduling is start traffic delay in 2 seconds and delay after transmission of 15 seconds. Jitter also measured in this test run. Trial number of test is 4 and trial duration is 100 seconds. Traffic load steps are starts at 100 and end at 10 each step is 10 for load units percentage. Frame size bytes are 512, 1024, 1280, 1518. It needs to be mentioned that Ethernet IEEE 802.3 standard defined the minimum Ethernet frame size as 64 bytes and the maximum as 1518 bytes. The maximum frame size was later increased to 1522 bytes to allow for VLAN tagging. The minimum size of an Ethernet frame that carries an ICMP packet is 74 bytes however this simulation only considered the four earlier mentioned frame sizes (bytes).
\subsection{Test Setup}
\begin{itemize}
\item Number of Nodes and Links
For this simulation there was 10 nodes and 18 links which two of the nodes were main packet traffic generator and receivers. Nodes and links are shown in Fig.~\ref{img:10Nodes}.

\item Various Generated Traffic Load
There are two sources and two destinations in this simulation which are both can be source or destination of packet traffic. Few nodes in the illustrated small subnetwork are also generating packets that are transmitting in between other nodes but not to the two nodes that are packet traffic generators/receivers. These extra packets generated by node 5 and node 3 at some interval of testing caused line rates exceed their capacity hence considerable oversubscription packets counts.The frame loss percentage demonstrated in Fig.~\ref{img:Frame Loss} and ~\ref{img:Frame_Loss}.
Monitoring the subnetwork for link congestion by our proposed algorithm via ACL packet filtering when load is more than pre-specified threshold could prevented such packet loss and oversubscription.
Our algorithm suggests setting a threshold value to limit maximum packet traffic allowed passing through the involved links in between nodes before packet traffic flow starts.

\item Frame Loss Test
This test was to determine the percentage of frames that should have been forwarded by the subnetwork node under steady state load that were not forwarded due to lack of resources. Result of frame loss of this test also considered the scenario of nodes 5 and 3 generated extra 74 bytes every 10 seconds interval and that was the reason that packet oversubscription happened on lower load percentage which was at 95 percent.Throughput result to determine the maximum rate at which none of the offered frames are dropped by the nodes under test along with frame loss test been run for the test setup as well.  
Result are shown in Fig.~\ref{img:Four Test Trials} and ~\ref{img:Test Results}.

\begin{figure}[ht]
\centering
\hspace*{-5mm}
\includegraphics[width=0.80\columnwidth]{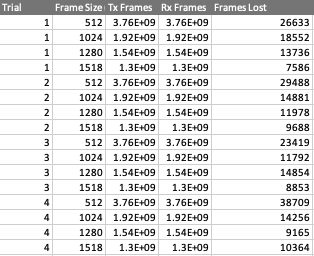}
\caption{Four Test Trials for Different Frame Sizes}
\label{img:Four Test Trials}
\end{figure}

\begin{figure}[ht]
\centering
\hspace*{-5mm}
\includegraphics[width=0.80\columnwidth]{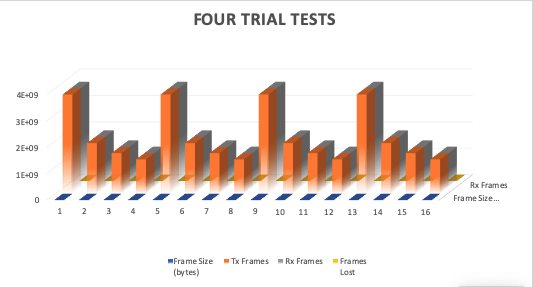}
\caption{Test Results}
\label{img:Test Results}
\end{figure}

\end{itemize}

\section{Summary}
This paper covered the problem of oversubscription packets due to traffic loads more than link line rates. Proposed algorithm suggest setting a threshold value for link loads less than their actual full capacity and take necessary actions when access control lists identify traffic loads above that threshold value.  Considering a control system which is real-time and state-dependent proposed access control list should lead into congestion control feature. This configurable value should allow network administrators to set policies and thresholds and specify how the system react when faced with a heavy packet traffic load situation. 
However, the drawback might be continues or large numbers of these condition within a specific time interval may have an impact on system’s ability to service subscriber sessions. Focusing need to be on real-time traffic in between network ports in order to identify links under heavy packet traffic load since most congestion occurrence in networks are temporarily. 
Future work for proposed idea is that including ACL rule of facilitating specific IP protocol in ACL stacks to become active when at times, targeted subnetworks face situations of heavy packet traffic loads.


\begin{thebibliography} {9}

\bibitem{farago} A. Faragó and Z. Ranjbar Mojaveri, ”In Search of the Densest Subgraph,” {\em Algorithms, Vol. 12(8),} 2019, pp. 157-175. 

\bibitem{waka} K. Wakabayashi, D. Kotani and Y. Okabe, "Traffic-aware Access Control List Reconstruction," {\em 2020 International Conference on Information Networking (ICOIN), Barcelona, Spain,} 2020, pp. 616-621. 

\bibitem{cisco} https://www.ciscopress.com/articles



\end{thebibliography}
\end{document}